\documentclass[%
 reprint,
superscriptaddress,
 amsmath,amssymb,
 aps,
]{revtex4-2}

\usepackage{graphicx}% Include figure files
\usepackage{dcolumn}% Align table columns on decimal point
\usepackage{bm}% bold math
\begin{document}

\preprint{APS/123-QED}

\title{Broadband Omni-resonant Coherent Perfect Absorption in Graphene}% Force line breaks with \\

\author{Ali K. Jahromi}
\email{jahromi@knights.ucf.edu}
\author{Massimo L. Villinger}
\author{Ahmed El Halawany}
\author{Soroush Shabahang}
\altaffiliation[Also at ]{Harvard Medical School and Wellman Center for Photomedicine, Massachusetts General Hospital, 50 Blossom Street, Boston, MA 02114, USA}
\author{H. Esat Kondakci}
\altaffiliation[Also at ]{Department of Physics, University of California, Santa Barbara, CA 93106, USA}
\affiliation{CREOL, The College of Optics \& Photonics, University of Central Florida, Orlando, FL 32816, USA}
\author{Joshua D. Perlstein}%
\affiliation{Department of Materials Science and Engineering, University of Central Florida, Orlando, FL 32816, USA}

\author{Ayman F. Abouraddy}
\affiliation{CREOL, The College of Optics \& Photonics, University of Central Florida, Orlando, FL 32816, USA}%

\date{\today}% It is always \today, today,

\begin{abstract}
Coherent perfect absorption (CPA) refers to interferometrically induced complete absorption of incident light by a partial absorber independently of its intrinsic absorption (which may be vanishingly small) or its thickness. CPA is typically realized in a resonant device, and thus cannot be achieved over a broad continuous spectrum, which thwarts its applicability to photodetectors and solar cells, for example. Here, we demonstrate broadband omni-resonant CPA by placing a thin weak absorber in a planar cavity and pre-conditioning the incident optical field by introducing judicious angular dispersion. We make use of monolayer graphene embedded in silica as the absorber and boost its optical absorption from $\approx1.6\%$ to $\sim60\%$ over a bandwidth of $\sim70$~nm in the visible. Crucially, an analytical model demonstrates that placement of the graphene monolayer at a peak in the cavity standing-wave field is \textit{not} necessary to achieve CPA, contrary to conventional wisdom.
\end{abstract}

\maketitle

%\tableofcontents
\section{Introduction}

Coherent perfect absorption (CPA) refers to the complete absorption of light incident on a partially absorbing structure \cite{Chong10PRL, Wan11Science} (see also early premonitions of this concept in \cite{Kaplan06OL}). The essence of CPA is that the interaction of an incoming optical field with a partial absorber can be maximized by engineering the field distribution \cite{Baranov17NRM}. Indeed, CPA suggests that by arranging for two fields to interfere in a lossy planar structure, one can \textit{always} guarantee that 100\% of the incident radiation is absorbed -- independently of the structure's intrinsic absorption or its thickness. This can be achieved by exploiting two counter-propagating fields incident on the two ports of a planar Fabry-P{\'e}rot (FP) resonator containing a partially absorbing layer between two mirrors M$_1$ and M$_2$ of reflectivities $R_{1}$ and $R_{2}$, respectively \cite{Wan11Science,Villinger15OL}. If the intrinsic single-pass absorption in a layer of an optical material is $\mathcal{A}$, then providing symmetric mirrors each having a reflectivity $R_{1}\!=\!R_{2}\!=\!1-\mathcal{A}$ guarantees that counter-propagating fields with prescribed relative amplitude and phase are completely absorbed \cite{Villinger15OL} -- even when $\mathcal{A}$ is vanishingly small. Alternatively, in the case of an ultrathin layer, placing it at a peak in a standing-wave resulting from the interference of two counter-propagating fields can yield complete absorption \cite{Roger15NatComm}.

A variety of applications in photonics can benefit from maximizing optical absorption in a given structure, ranging from optical detectors \cite{Engeln98RSI, Fiedler03CPL} to solar cells \cite{Pillaia07JAP, Rand04JAP}. Furthermore, a broad range of configurations can exploit the large absorption modulation achievable in CPA, including optical logic gates and switches \cite{Papaioannou16APLPhotonics, Rao14OL, Roger15NatComm, Rothenberg16OL, Fannin16IEEEPJ, Zhao16PRL, Papaioannou16LSA, Fang14APL}. Despite the exciting prospects afforded by CPA, several hurdles need to be first overcome, especially in applications where \textit{broadband} absorption is desired. First, optical absorption in any medium over a finite bandwidth is wavelength dependent $\mathcal{A}\!=\!\mathcal{A}(\lambda)$. As such, the cavity mirrors are required to have a wavelength-dependent reflectivity $R_{1}(\lambda)\!=\!R_{2}(\lambda)\!=\!1-\mathcal{A}(\lambda)$ that counterbalances the dispersive absorption, which can be realized via \textit{aperiodic} multilayer dielectric mirrors \cite{Pye17OL,Shabahang20Arxiv}. Therefore, CPA is in principle materials-agnostic.

Second, practical settings in optical detection or solar cells necessitate \textit{single}-port incidence rather than a counter-propagating-field configuration. This can be accommodated by placing the absorbing layer in an \textit{asymmetric} cavity having a back-reflector of reflectivity $R_{2}\!=\!1$ \cite{Furchi12NanoLett,Villinger15OL,Zhu16LSA,Pye17OL}. In this case, light is incident on a front mirror that must have a reflectivity $R_{1}(\lambda)\!=\!(1-\mathcal{A}(\lambda))^{2}$ to realize complete absorption \cite{Villinger15OL,Pye17OL}. Such a configuration is of course reminiscent of critical coupling in micro-cavities \cite{Yariv00EL, Cai00PRL, Yariv02PTL}. Eliminating the need for counter-propagating fields in turn eliminates the restrictions on the relative amplitude and phase, thus making the arrangement amenable to broadband incoherent light. Indeed, we recently demonstrated CPA in an asymmetric cavity over a bandwidth of $\sim\!60$~nm utilizing a fluorescent dye \cite{Shabahang20Arxiv}, and over a full octave of bandwidth $\sim\!800$-1600~nm using 2-$\mu$m-thick poly-crystallince silicon layer \cite{Pye17OL}.

The third and most serious hurdle is the resonant nature of this CPA configuration. For example, a FP resonator can satisfy the CPA condition over extremely broad bandwidths, but absorption is nevertheless harnessed only at the discrete resonant wavelengths within this spectrum \cite{Pye17OL, Jahromi18IEEEPJ}. This is clearly inadequate for enhancing the performance of a photodetector or solar cell, where absorption over a broad \textit{continuous} bandwidth is imperative. We have recently developed the concept of `omni-resonance', which can help yield continuous-wavelength CPA in a FP resonator. By introducing a specific angular dispersion into the field such that each wavelength is assigned to a prescribed angle, the entire incident spectrum -- which can significantly exceed the cavity free spectral range -- resonates with a single longitudinal cavity mode. The bandwidth of the resonant spectrum is thus decoupled from the cavity photon lifetime and quality factor. Combining CPA with omni-resonance in a FP cavity containing a partially absorbing layer can thus yield resonantly enhanced absorption over a broad, continuous spectrum.

In this paper, we demonstrate that broadband CPA can be realized in an ultrathin weakly absorbing layer (for instance, 2D materials such as graphene) once embedded in the appropriate \textit{planar} photonic environment. Besides their remarkable electronic transport features\cite{Geim07NatMater}, 2D materials strongly couple to light, suggesting them as intriguing candidates for optoelectronics\cite{Bonaccorso10NatPhoton}. Indeed, monolayer graphene has an intrinsic single-pass absorption in free space of $\mathcal{A}_\mathrm{G}\!\approx\!2.3\%$ despite being atomically thin \cite{Nair08Science}, which can make it potentially visible even to the naked eye. Nevertheless, $\mathcal{A}_\mathrm{G}$ is too low for most optoelectronics applications, including optical modulators and photovoltaics. Recent strategies for enhancing optical absorption in graphene include exploiting the extended interaction lengths afforded by waveguides \cite{Liu11Nature, Li12APL}, tuning the absorption via an electrostatic gate (also know as `chemical doping') \cite{Ju11NatNano, Fang13ACSNano}, placing graphene in a standing wave formed by counterpropagating fields \cite{Rao14OL,Roger16PRL}, or resonantly enhancing the absorption via guided-mode resonances \cite{Fannin16IEEEPJ, Liu14APL, Grande14OE, Grande15OptExp}, attenuated total internal reflection \cite{Zhao13OL, Pirruccio13ACSNano}, Salisbury screens \cite{Thareja15NanoLett, Jang14NanoLett, Ying17JAP, Woo14APL, Kakenov16ACSPhotonics}, or critical coupling \cite{Furchi12NanoLett, Zhu16LSA, Yao14NanoLett}.  \cite{Pirruccio13ACSNano}

\begin{figure*}[t!]
\centering\includegraphics[width=17.6cm]{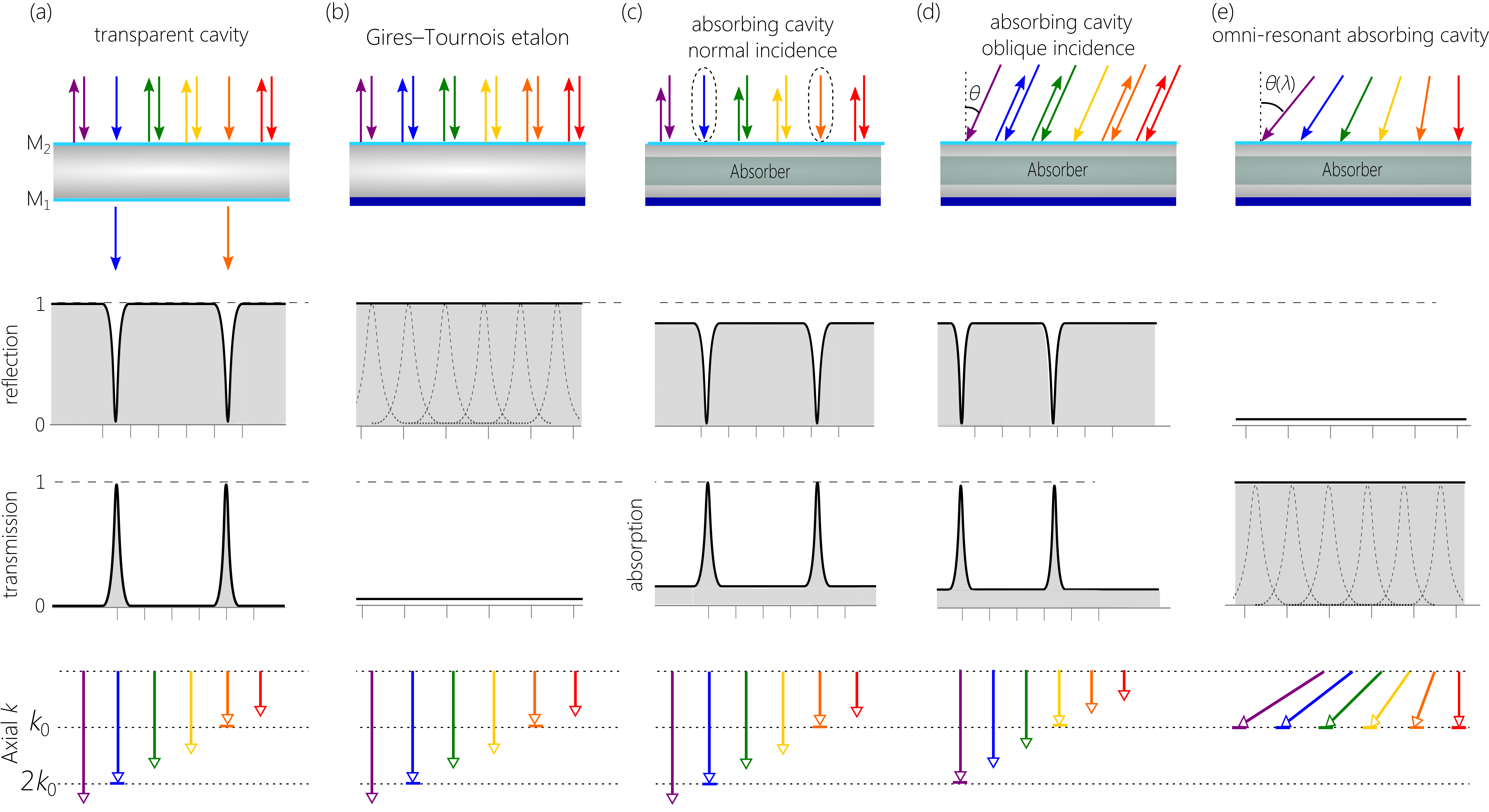}
\caption{Achieving broadband, spectrally continuous, omni-resonant CPA in a planar FP resonator by engineering the angular dispersion introduced into the incident field. (a) A collimated broadband field is normally incident on a passive symmetric FP cavity (first row) that transmits light at resonant wavelengths and reflects light otherwise (second and third rows). The resonant wavelengths are those for which the cavity wave vector is an integer multiple of $k_{0}\!=\!\pi/L$ (fourth row; ignoring the mirror reflection phases). (b) Same as (a) except that the passive FP cavity is \textit{asymmetric}: it is provided with a back-reflector $R_{2}\!=\!1$ (first row); the cavity therefore exhibits a spectrally flat, unity reflectivity (second row) and zero transmission (third row). The resonant wavelengths are the same as those for the symmetric cavity in (a). (c) After inserting a low-absorption material of single-pass absorption $\mathcal{A}$ (first row) into the cavity in (b), complete absorption is guaranteed when $R_{2}\!=\!1$ and $R_{1}\!=\!(1-\mathcal{A})^{2}$, and CPA is thus realized at the resonant wavelengths, resulting in spectral reflection dips (second row) and absorption peaks (third row). (d) At oblique incidence, the resonances in (c) are blue-shifted. (e) When the incident field is pre-conditioned such that each wavelength $\lambda$ is assigned to an appropriate incidence angle $\theta(\lambda)$ (first row), the wavelengths resonate continuously with the cavity; reflection over the full spectrum is eliminated (second row) and broadband CPA is achieved (third row); and the longitudinal wave-vector components of the incident wavelengths correspond to a \textit{single} resonant cavity mode (fourth row).}
\label{Fig:Concept}
\end{figure*}

Here, we demonstrate resonantly enhanced absorption ($\sim\! 20\!-\!60\%$) in a monolayer of graphene placed within a FP cavity continuously across a bandwidth of $\approx\!70$~nm in the visible. The cavity is designed to realize CPA at visible wavelengths and omni-resonance is simultanelously achieved by pre-conditioning the incident broadband incoherent field. Consequently, resonantly enhanced absorption is delivered over a broad, continuous spectrum. We therefore establish a general framework for achieving broadband perfect absorption in high-quality-factor (narrow-linewidth) optical cavities. Our motivation for selecting monolayer graphene is that it is a model for ultrathin absorbers with low intrinsic single-pass absorption. Furthermore, because its absorption is spectrally flat, we can focus on the juxtaposition of CPA with omni-resonance without the complications arising from wavelength-dependent absorption (which we dealt with in \cite{Villinger15OL,Pye17OL}). Furthermore, graphene has been shown to be useful in a variety of device architectures \cite{Bonaccorso10NatPhoton} that can benefit from maximizing absorption \cite{Zheng14ACSPhotonics}, particularly when realizing strong coupling between excitons and cavity photons \cite{Liu15NatPhoton, Zhao20OL}.

Realizing CPA by placing an ultrathin absorber such as graphene in a planar FP cavity presents a new configuration that has not been carefully examined thus far. Previous studies of CPA in planar cavities presumed `distributed' absorption over a finite thickness rather than `localized' absorption in a deeply subwavelength absorber \cite{Villinger15OL, Pye17OL}. Alternatively, absorption enhancement in an ultrathin absorber has been realized by placing it at a particular position within a standing wave formed of counterpropagating fields \cite{Altuzarra17ACSPhotonics, Rao14OL, Roger16PRL, Roger15NatComm, Papaioannou16APLPhotonics, Papaioannou16LSA}. It is commonly understood that maximizing the absorption necessitates placing the absorber at a standing-wave peak. This apparently sets stringent tolerances upon the position of the absorber within a FP cavity to maximize the absorption because the standing wave cannot be adjusted by modifying a field incident on a single port. We show here that the CPA-cavity setting surprisingly provides more favorable conditions for resonant absorption enhancement compared to the free-space configuration. Indeed, we find that maximum absorption can be achieved even if the absorber is \textit{not} located at a standing-wave peak, which \textit{cannot} be achieved in free space. Furthermore, graphene and other 2D materials offer a unique opportunity for post-fabrication tuning of the resonant absorption by adjusting the intrinsic absorption through an electrical potential.

This paper is organized as follows: first, we briefly review the concept of omni-resonance in the context of symmetric transparent FP cavities \cite{Shabahang17SciRep} and asymmetric FP cavities incorporating a partially absorbing layer \cite{Villinger20arxiv}. Next, we describe theoretically the distinction between two scenarios of `distributed' and `localized' absorption. We then present our experimental results on CPA in a graphene monolayer embedded in a planar FP cavity in absence then in presence of omni-resonance, thereby confirming omni-resonance-driven continuous broadband CPA in the visible.

\section{Concept of omni-resonance}

The key to harnessing the resonantly enhanced optical absorption associated with CPA over a \textit{broad}, \textit{continuous} bandwidth -- rather than only at discrete resonant wavelengths -- is the notion of omniresonance \cite{Shabahang17SciRep,Shabahang19OL,Shiri20OL,Shiri20APLP,Villinger20arxiv}, which is illustrated in Fig.~\ref{Fig:Concept}. Consider first a lossless symmetric planar FP cavity comprising a transparent layer of thickness $L$ and refractive index $n$ between a front mirror M$_1$ and a back mirror M$_2$ having equal reflectivities $R_{1}\!=\!R_{2}\!=\!R$; Fig.~\ref{Fig:Concept}(a). At normal incidence, light is transmitted only at the cavity resonances, whereupon the roundtrip phase $\varphi$ satisfies the condition \cite{Saleh07Book}:
\begin{equation}\label{Eq:NormalResonanceCondition}
\varphi(\lambda)=4\pi n\frac{d}{\lambda}+\alpha_{1}(\lambda)+\alpha_{2}(\lambda)=2m\pi,
\end{equation}
where $m$ is an integer, and $\alpha_{1}$ and $\alpha_{2}$ are the reflection phases from M$_1$ and M$_2$, respectively. Of course, if M$_2$ is a back-reflector having $R_{2}\!=\!1$ (i.e., a Gires-Tournois etalon \cite{Kuhl86JQE}), then the entire spectrum is reflected; Fig.~\ref{Fig:Concept}(b). Introducing an absorbing layer of single-pass absorption $\mathcal{A}(\lambda)$ into this cavity changes its response notably; Fig.~\ref{Fig:Concept}(c). Indeed, if $R_{1}(\lambda)\!=\!(1\!-\!\mathcal{A}(\lambda))^{2}$, then light is fully absorbed on resonance, $\mathcal{A}_{\mathrm{tot}}(\lambda)\!=\!1$, whenever $\lambda$ satisfies Eq.~\ref{Eq:NormalResonanceCondition} -- even when $\mathcal{A}$ is vanishingly small \cite{Villinger15OL,Pye17OL}. This is a manifestation of CPA, which is analogous here to critical coupling \cite{Yariv00EL,Cai00PRL,Yariv02PTL}.

The basis for the omni-resonance strategy can be appreciated by examining oblique incidence at an angle $\theta$ with respect to the cavity normal; Fig.~\ref{Fig:Concept}(d). The oblique-incidence resonant condition is determined by the \textit{longitudinal} component of the cavity wave vector:
\begin{equation}
\varphi(\lambda,\theta)=4\pi n\frac{d}{\lambda}\cos{\theta'}+\alpha_{1}(\lambda,\theta')+\alpha_{2}(\lambda,\theta')=2m\pi,
\end{equation}
where $\alpha_{1}(\lambda,\theta')$ and $\alpha_{2}(\lambda,\theta')$ are the reflection phases from M$_1$ and M$_2$, respectively, at a free-space wavelength $\lambda$ and an internal angle of incidence $\theta'$, which is related to the external angle of incidence $\theta$ through Snell's law: $\sin{\theta}\!=\!n\sin{\theta'}$ (assuming incidence from free space). As a result, all resonances \textit{blue-shift}: if $\lambda$ corresponds to a particular resonance at normal incidence, then the blue-shifted resonant wavelength $\lambda(\theta)$ at oblique incidence is
\begin{equation}
\lambda(\theta)=\frac{n}{\sqrt{n^{2}-\sin^{2}{\theta}}}\lambda,
\end{equation}
assuming that $\alpha_{1}(\lambda,\theta')$ and $\alpha_{2}(\lambda,\theta')$ are independent of $\theta'$ in the spectral range of interest, which is typically the case for dielectric multilayer Bragg mirrors within their bandgap.

Omni-resonance relies on arranging for \textit{each} wavelength $\lambda$ to impinge on the cavity at the appropriate angle $\theta(\lambda)$, such that it satisfies the omni-resonance condition:
\begin{equation}
\varphi(\lambda)\!=\!4\pi n\frac{d}{\lambda}\cos{(\theta'(\lambda))}+\alpha_{1}(\lambda,\theta'(\lambda))+\alpha_{2}(\lambda,\theta'(\lambda))\!=\!2m\pi.
\end{equation}
If this condition is satisfied over a certain bandwidth, it is guaranteed that the incident spectrum will resonate continuously with the cavity. Additionally, if the omni-resonance condition is met in conjunction with $R_{2}\!=\!1$ and the proper design of $R_{1}(\lambda)$, then the entire spectrum is absorbed and broadband omni-resonant CPA is achieved; Fig.~\ref{Fig:Concept}(e). This overall strategy is materials-agnostic, and is independent of the physical nature of the absorbing layer. Furthermore, the design can be adjusted in principle for any desired central wavelength or bandwidth. The central challenge is to construct an optical pre-conditioning system that introduces the particular wavelength-dependent angle of incidence $\theta(\lambda)$ that satisfies the omni-resonance condition, which we present below.

\section{Impact of graphene location inside the cavity}
\label{Sec:GrapheneLoc}

Our previous work on CPA using the configuration shown in Fig.~\ref{Fig:Concept}(c) assumed `distributed' absorption; that is, the single-pass absorption $\mathcal{A}$ results from a layer in which absorption extends uniformly over its thickness $L$ \cite{Villinger15OL,Pye17OL,Villinger20arxiv,Shabahang20Arxiv}. This assumption no longer holds for `localized' absorption (such as a 2D material like graphene or a metasurface). It is expected that the location of a deeply subwavelength-thick absorber with respect to the peaks and minima of the cavity standing-wave field structure will impact the net absorption. Intuitively, light will not interact with the absorber when placed at a standing-wave null, whereas placing it at a peak will likely maximize absorption $\mathcal{A}_{\mathrm{tot}}$. Indeed, previous work on CPA in ultrathin layers has indicated that maximum absorption is achieved \textit{only} when the layer is placed at a standing-wave peak \cite{Rao14OL, Papaioannou16APLPhotonics}. 

\begin{figure}[t!]
\centering\includegraphics[width=8.6cm]{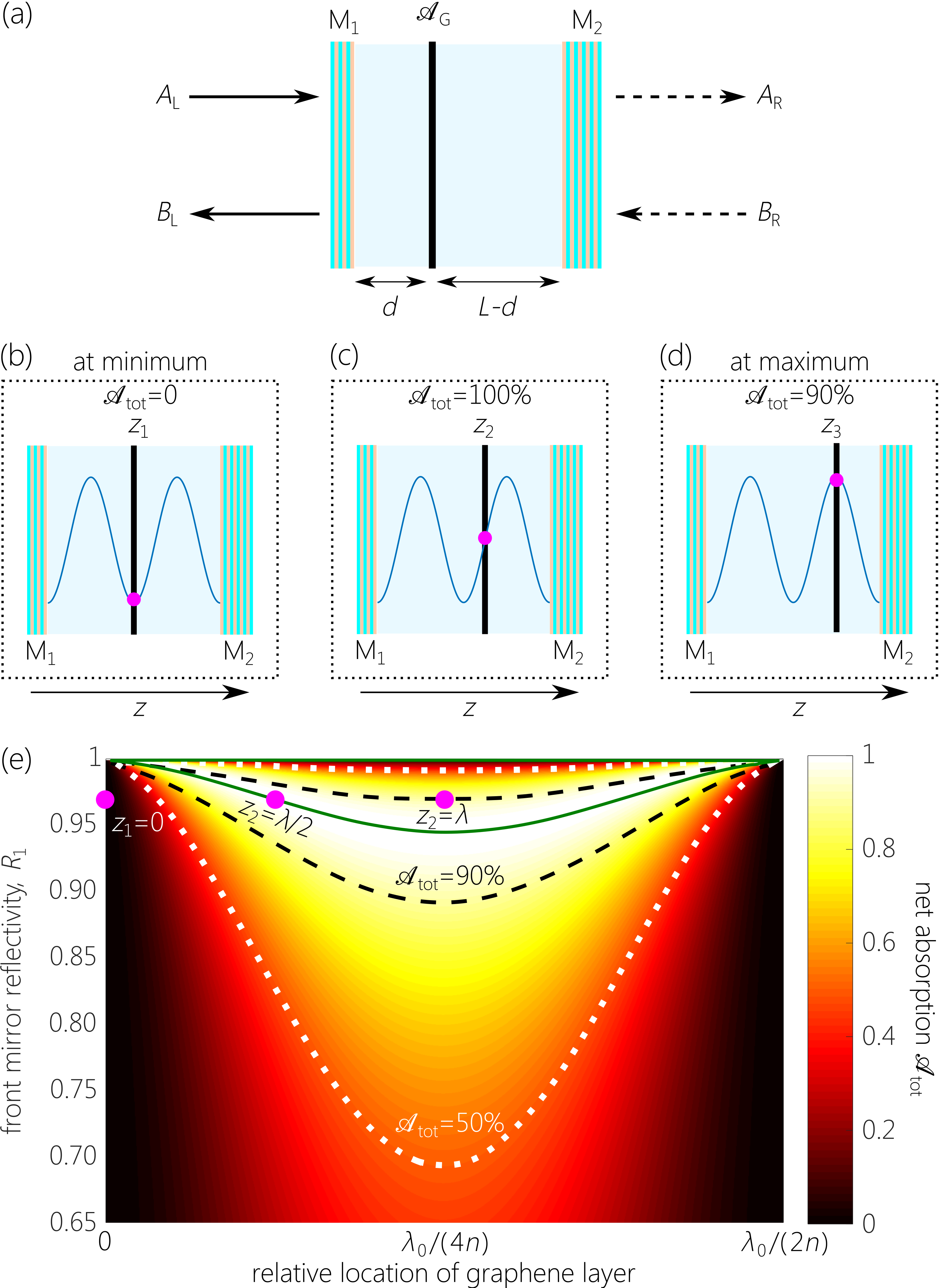}
\caption{Impact of the position of an ultrathin absorbing layer on the net cavity absorption $\mathcal{A}_{\mathrm{tot}}$. (a) Schematic depiction of a planar FP resonator containing a localized absorber having single-pass absorption $\mathcal{A}_{\mathrm{G}}$. (b-d) Three scenarios in a FP cavity having $R_1\!=\!(1-\mathcal{A}_\mathrm{G})^2\!\approx\!96.8\%$, which corresponds to the CPA condition for a distributed absorber with $\mathcal{A}\!=\!\mathcal{A}_{\mathrm{tot}}$. The thin absorber, which is represented by a thick black line, is considered at three distinct locations within the cavity standing-wave. (b) When the absorber is placed at the standing-wave null $z_1\!=\!0$, $\mathcal{A}_{\mathrm{tot}}\!=\!0$. (c) Placing the absorber midway between a null and peak $z_2\!=\!\lambda'/8$ results in perfect absorption $\mathcal{A}_{\mathrm{tot}}\!=\!1$; $\lambda'$ is the wavelength inside the cavity. (d) When the absorber is placed at a standing-wave peak $z_3\!=\!\lambda'/4$, the net absorption is sub-optimal $\mathcal{A}_\mathrm{tot}\!\approx\!0.9$. (e) Calculated $\mathcal{A}_{\mathrm{tot}}$ as a function of the absorbing layer position $z$ and the front-mirror reflectivity $R_{1}$. The three red dots correspond to the positions $z_{1}$, $z_{2}$, and $z_{3}$ in (b), (c) and (d), respectively. The solid curve is the CPA contour $\mathcal{A}_\mathrm{tot}\!=\!1$. The dashed curves are the contours for $\mathcal{A}_\mathrm{tot}\!\approx\!0.9$, and the dotted curves are for $\mathcal{A}_\mathrm{tot}\!\approx\!0.5$.} 
\label{Fig:PositionTheory}
\end{figure}

In this Section we evaluate the impact of the \textit{position} of an ultrathin absorbing layer within the cavity on $\mathcal{A}_{\mathrm{tot}}$ and we demonstrate that the cavity shown schematically in Fig.~\ref{Fig:PositionTheory}(a) modifies previous predictions in a fundamental way. The incident and reflected field amplitudes on the left at normal incidence are $A_{\mathrm{L}}$ and $B_{\mathrm{L}}$, respectively, the corresponding quantities on the right are $A_{\mathrm{R}}\!=\!0$ (the field is incident from the left only) and $B_{\mathrm{R}}$, which are related through the cavity transfer matrix $\mathbf{M}$,
\begin{equation}
\begin{pmatrix} A_\mathrm{L}\\B_\mathrm{L}\end{pmatrix}=\mathbf{M}\begin{pmatrix}A_\mathrm{R}\\B_\mathrm{R}\end{pmatrix}.
\end{equation}
The cavity length is $L$, and the absorber is placed at a distance $d$ from mirror M$_1$. Therefore, from left to right, the cavity comprises mirror M$_1$, propagation a distance $d$, the absorbing layer, propagation a distance $L-d$, and mirror M$_2$, cascaded according to :
\begin{equation}
\mathbf{M}=\mathbf{M}_{1}\cdot\mathbf{P}(\varphi_{1})\cdot\mathbf{M}_\mathrm{G}\cdot\mathbf{P}(\varphi_{2})\cdot\mathbf{M}_{2};
\end{equation}
here $\mathbf{M}_{1}$ and $\mathbf{M}_{2}$ represent mirrors M$_1$ and M$_2$, respectively, which have the form given in Eq.~\ref{Eq:MirrorTransferMatrix}; $\mathbf{P}(\varphi_{1})$ and $\mathbf{P}(\varphi_{2})$ represent propagation for distances $d$ and $L-d$, respectively, where $\varphi_{1}\!=\!nkd$, $\varphi_{2}\!=\!nk(L-d)$, $k=\tfrac{2\pi}{\lambda}$, $\lambda$ is the free-space wavelength, and $\mathbf{P}(\varphi)$ has the form given in Eq.~\ref{Eq:PropTransferMatrix}; and $\mathbf{M}_\mathrm{G}$ represents the ultrathin absorbing layer, taken to be monolayer graphene for concreteness \cite{Zhan13JPCM}: 
\begin{equation}
\mathbf{M}_{\mathrm{G}}\!=\!\begin{pmatrix} 1\!+\!\delta & \delta \\ -\delta & 1\!-\!\delta \end{pmatrix},
\end{equation}
where $\delta\!=\!\tfrac{\sigma\mu_{\mathrm{o}}c}{2n}$, $\mu_{\mathrm{o}}$ is the free-space magnetic permeability, $c$ is the light speed in vacuum, $n$ is the refractive index of the surrounding medium, and the optical conductivity of monolayer graphene is $\sigma\!=\!6.1\!\times\!10^{-5}$~$\Omega^{-1}$ \cite{Mak08PRL}. Accordingly, the single-pass absorption is $\mathcal{A}_\mathrm{G}\!=\!\frac{2\delta}{1+\delta^2}\!\approx\!2\delta\!=\!1.6\%$ ($\delta\!\ll\!1$).

The cavity reflection amplitude $r_\mathrm{L}\!=\!\tfrac{B_{\mathrm{L}}}{A_{\mathrm{L}}}$ is given by:
\begin{equation}
r_{\mathrm{L}}\!=\!-e^{i(2\beta_{1}-\alpha_{1})}
\frac
{
r_{1}\delta_{+}e^{-i\psi_{\mathrm{o}}}
-\delta_{-}e^{i\psi_{\mathrm{o}}}
+\delta e^{-i\Delta\psi}
+r_{1}\delta e^{i\Delta\psi}}
{
\delta_{+}e^{-i\psi_{\mathrm{o}}}
-r_{1}\delta_{-}e^{i\psi_\mathrm{o}}
+r_{1}\delta e^{-i\Delta\psi}
+\delta e^{i\Delta\psi}
},
\end{equation}
where $\delta_{+}\!=\!1\!+\!\delta$, $\delta_{-}\!=\!1\!-\!\delta$, $\varphi_{\mathrm{o}}\!=\!nkL\!=\!\varphi_{1}\!+\!\varphi_{2}$, $2\psi_{\mathrm{o}}\!=\!2\varphi_{\mathrm{o}}\!+\!\alpha_{1}\!+\alpha_{2}$ is the cavity round-trip phase, $\Delta\psi\!=\!\Delta\varphi+\tfrac{\alpha_{2}-\alpha_{1}}{2}$, $\Delta\varphi\!=\!\varphi_{2}\!-\!\varphi_{1}$, $\beta_{1}$ and $\alpha_{1}$ are the transmission and reflection phases for M$_1$, respectively, $\alpha_{2}$ is the reflection phase for M$_2$, and $r_{1}$ is the reflection amplitude from M$_1$ ($R_{1}\!=\!|r_{1}|^{2}$) for incidence from within the cavity (from a medium of refractive index $n$). The net cavity absorption is $\mathcal{A}_{\mathrm{tot}}\!=\!1-|r_{\mathrm{L}}|^{2}$, and we maximize the resonant absorption by finding $r_{1}$ that minimizes $|r_\mathrm{L}|$. On resonance $\psi_{\mathrm{o}}\!=\!m\pi$, by setting $\Re\{r_\mathrm{L}\}\!=\!0$ ($\Im\{r_\mathrm{L}\}\!\ll\!1$) we find the optimal $R_{1}$ to realize $\mathcal{A}_{\mathrm{tot}}\!\approx\!1$ is:
%\begin{eqnarray}
%\label{eqn:Opt_r1}
%R_{1}\approx\left(1-2\mathcal{A}_\mathrm{G}\sin^2(\Delta\psi/2)\right)^{2}.
%\end{eqnarray}

\begin{eqnarray}
\label{eqn:Opt_r1}
R_{1}\approx\left(1-2\mathcal{A}_\mathrm{G}\sin^2\Big(\frac{m\pi d}{L}-\frac{\alpha_1+\alpha_2}{4}\Big)\right)^{2}.
\end{eqnarray}

\begin{figure}[t!]
\centering\includegraphics[width=8.6cm]{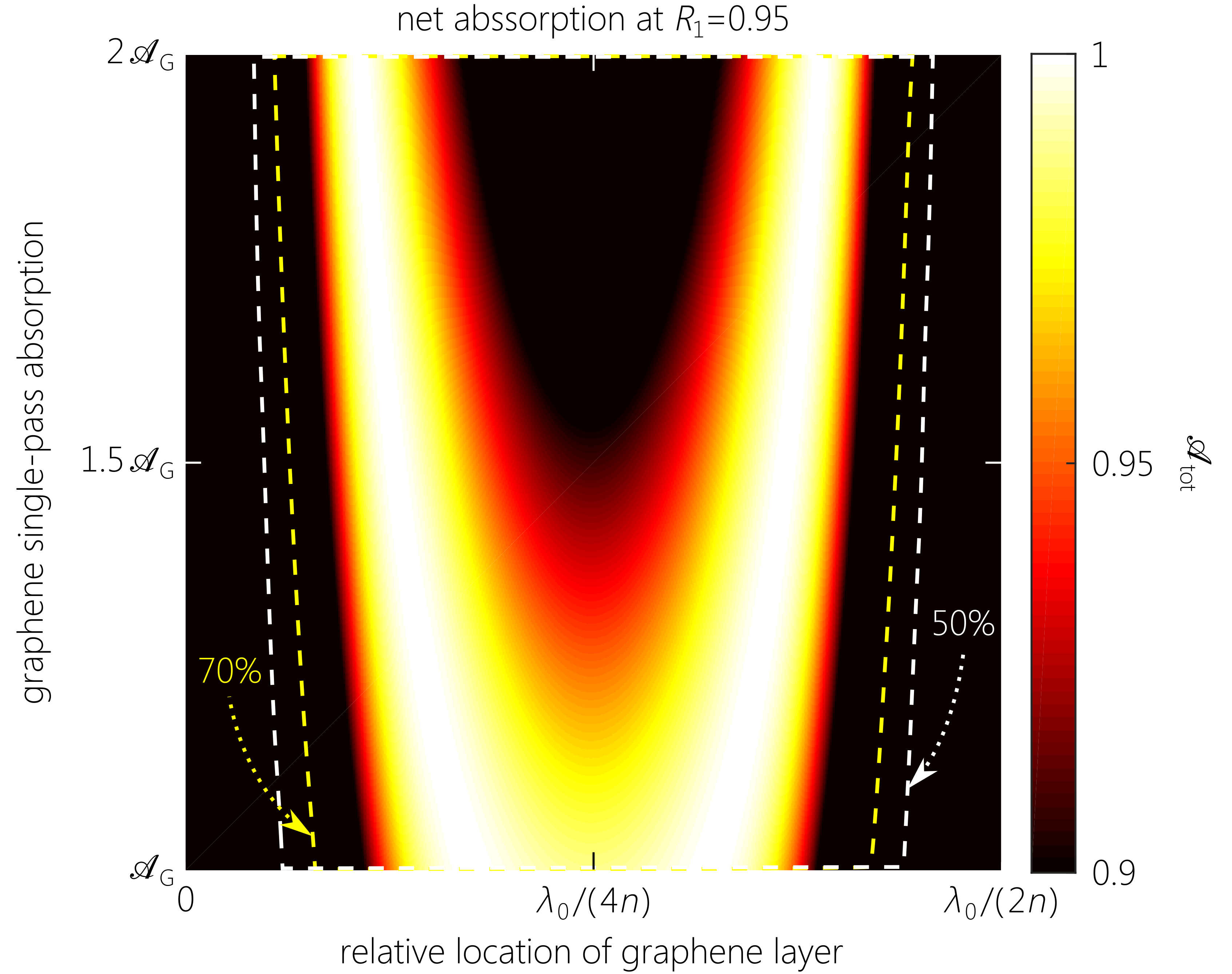}
\caption{Calculated cavity net absorption $\mathcal{A}_\mathrm{tot}$ as of function of the graphene position $z$ and the graphene single-pass absorption $\mathcal{A}$, for a given mirror reflectivity, $R_1\!=\!0.95$. Compared to Fig. \ref{Fig:PositionTheory}(e), this is more applicable plot since the mirror reflectivities are not changeable after the device fabrication, while the graphene absorption can be still varied in a range via electrical injection. Depending on the graphene position, the graph shows the cavity absorption can be quite modulated via tuning the graphene single-pass absorption. The white and the yellow dashed curves determine the areas within which the cavity absorption is higher than 50\% and 70\%, respectively.} 
\label{Fig:PositionTheory2}
\end{figure}

\begin{figure*}[t!]
\centering\includegraphics[width=17.6cm]{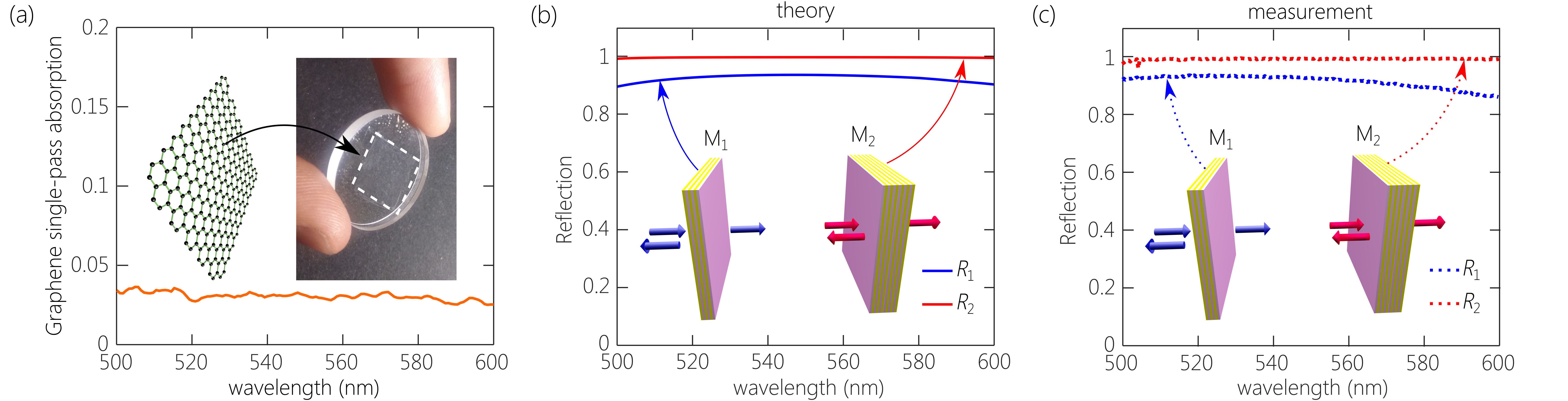}
\caption{(a) Measured single-pass-absorption spectrum $\mathcal{A}(\lambda)$ for a graphene monolayer. The inset is a photograph of a glass substrate on which a graphene layer is deposited. The graphene layer is delimited by dashed white square. (b) The calculated reflectivities for incidence from free space onto the back and front mirror M$_1$ and M$_2$, respectively, via the transfer matrix method in the 500-600~nm spectral range. Here M$_1$ and M$_2$ are constructed of 5 and 10 dielectric bilayers, respectively. (c) Measured spectral reflectivities for the fabricated mirrors M$_1$ and M$_2$.}
\label{Fig:CavityDesign}
\end{figure*}

We can now establish a connection between the location of the absorber and the total absorption by noting that the intensity of the cavity standing-wave in absence of the absorber is $I(d)\!\propto\!\sin^2(\Delta\psi/2)$. The phase $\Delta\psi$ therefore determines the location of the thin absorber with respect to the cavity standing-wave: $\Delta\psi\!=\!0$ corresponds to a null and $\Delta\psi\!=\!m\pi$ to a peak. When the absorber is placed at a null, Eq.~\ref{eqn:Opt_r1} indicates that the optimal reflectivity is $R_{1}\!\rightarrow\!1$, and absorption is eliminated, as one expects; Fig.~\ref{Fig:PositionTheory}(b). Surprisingly, however, the absorber need \textit{not} be placed at a peak to realize CPA. Indeed, if the absorber is placed anywhere except for the null $\Delta\psi\!\neq\!0$, then $\mathcal{A}_{\mathrm{tot}}\!\approx\!1$ can always be maximized by selecting the reflectivity of M$_1$ according to Eq.~\ref{eqn:Opt_r1}. For example, if the absorber is placed midway between a null and a peak $\Delta\psi\!=\!\tfrac{\pi}{2}$, then selecting $R_{1}\!=\!(1-\mathcal{A}_{\mathrm{G}})^{2}$ yields complete absorption; Fig.~\ref{Fig:PositionTheory}(c). Note that this is the optimal reflectivity for M$_1$ when employing a \textit{distributed} absorber having single-pass absorption $\mathcal{A}_{\mathrm{G}}$. If one makes use of this same mirror, but now places the absorber at a peak in the standing wave, as shown in Fig.~\ref{Fig:PositionTheory}(d), then maximum absorption is \textit{not} attained; indeed, here $\mathcal{A}_{\mathrm{tot}}\!=\!90\%$. Placing the absorber at this position requires instead a reflectivity $R_{1}\!=\!(1-2\mathcal{A}_{\mathrm{G}})^{2}$ to achieve maximum absorption.

The presence of a cavity therefore modifies the dependence of absorption on the location of the absorber in a fundamental way. Specifically, we find -- counter-intuitively -- that complete absorption can be reached for any position of the thin absorber (as long as it is \textit{not} at a null) if the appropriate reflectivity for M$_1$ is utilized. This is emphasized in Fig.~\ref{Fig:PositionTheory}(e) where we plot $\mathcal{A}_{\mathrm{tot}}$ as a function of the position $z$ of the absorber over an optical wavelength ($\Delta\psi\!=\!2\pi z/\lambda$) and $R_{1}$. From Fig.~\ref{Fig:PositionTheory}(e) it is clear that substantial tolerance is afforded by the cavity design ($R_{1}$ and the absorber position) to approach $\mathcal{A}_{\mathrm{tot}}\!\approx\!100\%$ for an ultrathin absorber. The region between the two dashed contours in the parameter space in Fig.~\ref{Fig:PositionTheory}(e) corresponds to $\mathcal{A}_{\mathrm{tot}}\!<\!90\%$ (an enhancement of $\approx\!55\times$ over $\mathcal{A}_{\mathrm{G}}$), whereas between the dotted contours $\mathcal{A}_{\mathrm{tot}}\!<\!50\%$ (an enhancement of $\approx\!30\times$). In other words, large resonant absorption enhancement can be achieved even when the absorber position deviates from the standing-wave peak, or $R_{1}$ does not match the targeted reflectivity. For example, taking $R_{1}\!=\!99.5\%$ results in $\mathcal{A}_{\mathrm{tot}}\!>\!50\%$ for $\sim\!90\%$ of the positions of the graphene monolayer within the cavity standing-wave. Indeed, Fig.~\ref{Fig:PositionTheory}(c,d) in which $R_{1}\!=\!96.8\%$ indicate that $\mathcal{A}_{\mathrm{tot}}\!>\!90$ is reached for $\sim\!50\%$ of the locations.

A more realistic scenario involves a FP cavity having a fixed $R_1$ and a graphene monolayer at a fixed position. In this case, by tuning $\mathcal{A}$ away from $\mathcal{A}_{\mathrm{G}}$ through chemical doping (by applying an electrical potential) one may maximize $\mathcal{A}_{\mathrm{tot}}$. Indeed, graphene and other 2D materials offer tremendous flexibility in this regard \cite{Kasry10ACSNano}. In Fig.~\ref{Fig:PositionTheory2} we plot $\mathcal{A}_{\mathrm{tot}}$ for $R_1\!=\!0.95$ while varying the position of the graphene monolayer and its single-pass absorption $\mathcal{A}$ from $\mathcal{A}_{\mathrm{G}}$ to $2\mathcal{A}_{\mathrm{G}}$, which is accessible for graphene. It is clear that tuning $\mathcal{A}$ in this range allows for increasing the resonant absorption if the absorber is not placed at the targeted position within the cavity. This all bodes well for constructing planar devices that resonantly enhance the absorption in 2D materials without exorbitant fabrication tolerances.

\section{Design and characterization}

We now proceed to our experiments validating resonantly enhanced absorption in a graphene monolayer. The FP cavity we utilize comprises two multilayer dielectric Bragg mirrors sandwiching a silica spacer in which a graphene monolayer is embedded. In this Section we describe the components of the CPA cavity and the performance of the cavity as a whole on resonance, and in the next Section we will combine it with a light pre-conditioning system that establishes omni-resonance over a broad continuous bandwidth.

Because the graphene's intrinsic single-pass spectral absorption $\mathcal{A}(\lambda)\!\approx\!\mathcal{A}_{\mathrm{G}}\!\approx\!1.6\%$ in silica is flat over a broad spectrum as shown in Fig.~\ref{Fig:CavityDesign}(a) \cite{Nair08Science}, the mirror reflectivity required to realize CPA at any wavelength within this range is also spectrally flat $R_{1}(\lambda)\!\approx\!R_{1}\!=\!(1-\mathcal{A}_{\mathrm{G}})^{2}\!\approx\!96.8\%$, which can be realized with periodic dielectric Bragg mirrors. The light source used in all our measurements is a halogen lamp (Thorlabs, QTH10/M), and the spectrum is recorded with an optical spectrum analyzer (OSA; Advantest Q8383). Light from the halogen lamp is spatially filtered by coupling it into a 1-m-long multimode, 50-$\mu$m-diameter optical fiber terminated with a fiber collimator. 

The targeted reflectivities for M$_1$ and M$_2$ are shown in Fig.~\ref{Fig:CavityDesign}(b): $R_{2}\!=\!1$ for M$_2$, and $R_{1}\!\approx\!96.8\%$ for M$_1$. This reflectivity is selected assuming the graphene monolayer is placed midway between the peak and null of the cavity standing-wave field distrbution. Here $R_{1}$ is the reflectivity of M$_1$ for incidence from the silica spacer within the cavity. In Fig.~\ref{Fig:CavityDesign}(b) we plot the corresponding reflectivity for incidence from air to facilitate comparison with measurements. Both M$_1$ and M$_2$ are constructed of bilayers of 94-nm-thick SiO$_2$ and 61-nm-thick Ti$_2$O$_3$ that form a broad reflection bandwidth $\sim\!100$~nm centered at $\sim\!550$~nm. The mirrors M$_1$ and M$_2$ are formed of 5 and 10 bilayers, respectively, and their measured normal-incidence reflectivities from free space are plotted in Fig.~\ref{Fig:CavityDesign}(c), displaying excellent agreement with their theoretical counterparts in Fig.~\ref{Fig:CavityDesign}(b).

\begin{figure}[t!]
\centering\includegraphics[width=8.6cm]{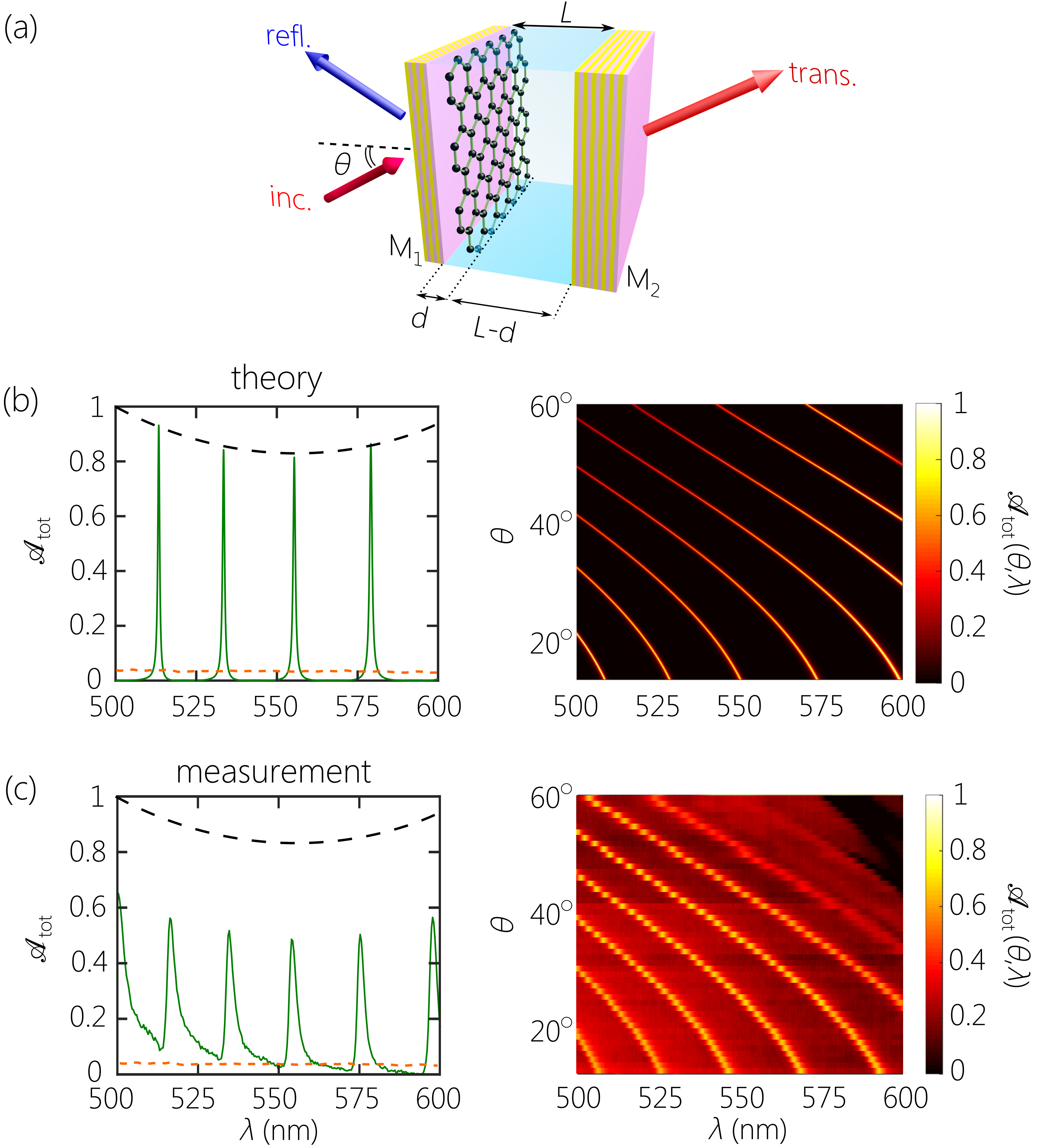}
\caption{Characterization of the CPA cavity containing the graphene monolayer. The is practically no transmitted field through the back-reflector M$_2$. (a) Schematic of the CPA. (b) Calculated net cavity spectral absorption for normally incident light is shown on the left, and as a function of the angle of incidence $\mathcal{A}_{\mathrm{tot}}(\lambda,\theta)$ on the right. (c) Measured absorption spectra corresponding to (b). The dashed curves in the left panels in (b) and (c) represent the theoretical limit on absorption given the experimentally realized value of the reflectivity $R_1$ for the front mirror M$_1$.}
\label{Fig:Characterization}
\end{figure}

The overall FP cavity structure for CPA is depicted schematically in Fig.~\ref{Fig:Characterization}(a). Its construction starts with depositing the back mirror M$_2$ onto a 1-mm-thick, 25-mm-diameter glass slide, followed by a 4-$\mu$m-thick silica spacer. The graphene monolayer (of nominal thickness 0.35~nm \cite{Bruna09APL}) is deposited on the spacer, and is then secured in place with a 100-nm-thick silica layer, before depositing the front mirror M$_1$. All deposition steps for the dielectric layers are carried out via e-beam evaporation (Appendix~\ref{App:MirrorDesign}). Using the transfer-matrix method, we calculate the net spectral absorption $\mathcal{A}_{\mathrm{tot}}\!=\!1-|r_{\mathrm{L}}|^{2}-|t_{\mathrm{L}}|^{2}$ (throughout we have $t_{\mathrm{L}}\!\approx\!0$) in the cavity as a function of the angle of incidence $\theta$. We plot $\mathcal{A}_{\mathrm{tot}}(\lambda,\theta)$ in Fig.~\ref{Fig:Characterization}(b) where it is clear that absorption enhancement is achieved at the cavity resonances. However, $100\%$ absorption is not reached here because of the finite steps in reflectivity achieved by adding each bilayer to M$_1$. Indeed, 4 bilayers yield $R_{1}\!=\!89\%$ for incidence from air and 5 bilayers yield $R_{1}\!=\!96\%$, whereas the target reflectivity is $R_{1}\!=\!94\%$ (for incidence from free space). The dashed curve in Fig.~\ref{Fig:Characterization}(b) is the theoretical limit on $\mathcal{A}_{\mathrm{tot}}$ given our particular cavity design. The corresponding measured absorption spectra are plotted in Fig.~\ref{Fig:Characterization}(c) for a broadband field of diameter $\approx\!10$~mm incident at angles in the range $0^{\circ}$ to $60^{\circ}$ with respect to the cavity normal. The absorption resonances do not reach the theoretical limit due to the imperfections in the deposited structure (departures from the targeted layer thicknesses and from an exact planar condition), and due to uncertainty in the exact location of the graphene monolayer within the silica spacer as discussed above. The absorption peaks extending over a bandwidth of 100~nm reach a maximum of $\approx\!60\%$.

\begin{figure}[t!]
\centering\includegraphics[width=8.6cm]{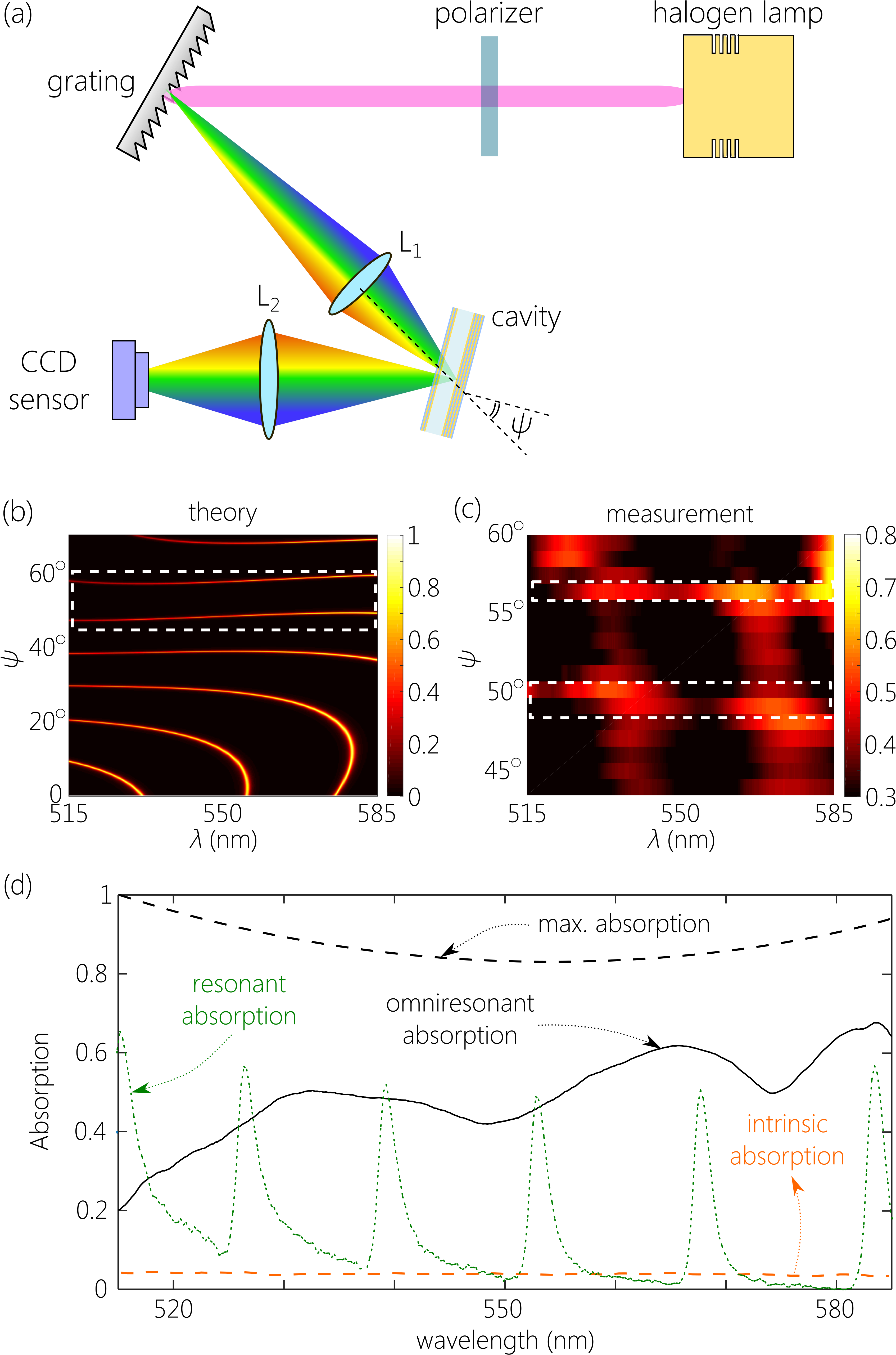}
\caption{Characterization of the CPA cavity in conjunction with omni-resonance. (a) Schematic of the pre-conditioning optical system to realize omni-resonance in the CPA cavity characterized in Fig.~\ref{Fig:Characterization}. (b) Calculation of the absorption spectrum as a function of the cavity tilt angle $\psi$. (c) Measured absorption spectra corresponding to theoretical plot in (b). The measurements were carried out  in increments of $1^{\circ}$ in the angular range of $40^{\circ}\!<\!\psi\!<\!70^{\circ}$, corresponding to the range highlighted with a white dashed rectangle in (b). The two cavity tilt angles $\psi\!=\!50^{\circ}$ and $\psi\!=\!57^{\circ}$ correspond to two distinct omni-resonances. (d) The omni-resonant spectral absorption at a cavity tilt angle $\psi\!=\!57^{\circ}$. For comparison, we also plot the single-pass absorption of the graphene monolayer, the theoretical limit on cavity absorption, and the resonant CPA in absence of omni-resonance.}
\label{Fig:Omniresonance}
\end{figure}

\section{Realization of broadband coherently enhanced omni-resonant absorption}

It is clear from Fig.~\ref{Fig:Characterization}(b,c) that the spectral absorption in the cavity at any incidence angle is resonantly enhanced only at discrete resonances rather than across a spectrally continuous band. The latter is realized by exploiting the concept of omni-resonance as illustrated in Fig.~\ref{Fig:Concept}(e). This requires introducing angular dispersion into the incident field to associate each wavelength $\lambda$ with a specific angle of incidence $\theta(\lambda)$ via an optical system that pre-conditions the incident field. The experimental arrangement for demonstrating omni-resonance is shown in Fig.~\ref{Fig:Omniresonance}(a). Broadband light from the halogen lamp is directed through a 1-mm-wide slit onto a reflective diffraction grating (Thorlabs GR25-1850; 1800~lines/mm, area $25\!\times\!25$~mm$^2$), which is oriented at $50^{\circ}$ with respect to the incident light. The spectrally resolved wave front in the first diffraction order is focused onto the cavity by a lens L$_1$, which is an aspheric condenser (Thorlabs, ACL50832U) placed at a distance 12~cm from the grating. The cavity is mounted on a rotational stage near the focal plane of L$_1$, and is tilted an angle $\psi$ with respect to an optical axis defined by a central wavelength $\lambda_{\mathrm{c}}\!=\!550$~nm. This combination of grating and lens L$_1$ provides the requisite linear angular dispersion over the wavelength range of interest \cite{Shabahang17SciRep}; see Appendix~\ref{App:SetupDesign} for further details. Light reflected from the cavity is collimated by a 25-mm-focal-length lens L$_2$ and directed to a monochrome CCD sensor (Thorlabs DCC1545M). After calibration with a tunable narrowband spectral filter ($\approx\!3$-nm bandwidth), the spatial positions on the CCD correspond to wavelengths, and the reflected spectrum is captured as an image on the sensor for each $\psi$.

The computed absorption spectrum as we vary the cavity tilt angle $\psi$ is plotted in Fig.~\ref{Fig:Omniresonance}(b). It is clear that at certain valuues of $\psi$ the absorption is spectrally flat; i.e., broadband omni-resonance has been achieved, and the attendant CPA is realized over a continuous broad spectrum. Examples of omni-resonant CPA occur at $\psi\!\approx\!50^{\circ}$ and $60^{\circ}$ in Fig.~\ref{Fig:Omniresonance}(b). Each of the angular settings for $\psi$ identifies an `achromatic resonance'. This is a single resonance of the bare cavity that is now spread over an extended, continuous spectral range rather than the narrow linewidth at a fixed wavelength as shown in Fig.~\ref{Fig:Characterization}(b). The re-orientation of the resonance trajectory in wavelength-angle $(\lambda.\psi)$ space is a consequence of the angular dispersion introduced into the incident field. Moreover, note that each achromatic resonance extends in bandwidth across multiple free spectral ranges of the bare cavity \cite{Shabahang17SciRep,Shabahang19OL,Villinger20arxiv}.

The measurement results for the spectral absorption while varying $\psi$ are plotted in Fig.~\ref{Fig:Omniresonance}(c). Because of the physical constraints in the measurement apparatus resulting from the size of the cavity and the focal lengths of the lenses L$_1$ and L$_2$, measurements were collected for cavity tilt angles over the range $45^{\circ}\!<\!\psi\!<\!60^{\circ}$. The measurements indicate that a prominent wideband absorption occurs at the tilt angle $\psi\!=\!57^\circ$ associated with an achromatic resonance. The absorption spectrum at this tilt angle is plotted in Fig. \ref{Fig:Omniresonance}(d), where it is compared to the corresponding spectra of the graphene monolayer and that of the bare cavity. We observe that the resonant enhancement in absorption is no longer confined to the resonant linewidths separated by the FSR. Rather, a continuous broadband spectrum of resonant enhancement is now realized as a result of the juxtaposition of CPA with omni-resonance. Two points must be noted here. First, the observed continuous absorption spectrum is \textit{not} a consequence of `filling in' the intra-resonance gaps. Instead, the absorption spectrum is associated with a \textit{single} achromatic resonance. Second, this achromatic resonance does \textit{not} correspond to any of the bare cavity resonances shown in Fig.~\ref{Fig:Characterization}(b,c). Instead, it corresponds to a higher-order bare-cavity resonance that becomes spectrally flat in our wavelength range of interest; see Fig.~\ref{Fig:Omniresonance}(b).

\section{Conclusion}

We have combined in the same device the physically independent phenomena of CPA (resonantly enhanced complete absorption in a planar cavity containing a partially absorbing layer) and omni-resonance (the continuous, broad resonance spectrum realized when light incident on the cavity is pre-conditioned by introducing the appropriate angular dispersion). By embedding a graphene monolayer of a single-pass absorption $\mathcal{A}_{\mathrm{G}}$ in a planar cavity provided with a back-reflector and an appropriately designed front mirror, light on resonance can be fully absorbed, despite the low intrinsic absorption of graphene. By adding a light pre-conditioning system, the resonantly enhanced absorption is now realized over a broad continuous spectrum and is associated with a single achromatic resonance. Our measurements reveal at one such achromatic resonance enhanced absorption to $\sim\!60\%$ in monolayer graphene observed over a continuous bandwidth of $\approx\!70$~nm.

We have also elucidated the impact of the position of an ultrathin absorber such as monolayer graphene on the net cavity absorption. We have shown that it is no longer necessary to place the absorbing layer at a peak in the standing-wave pattern within the cavity as is commonly understood to be the case in free space. Instead, maximum absorption is guaranteed to reach its maximum value if the absorber is placed anywhere except at a field null. This requires only adjusting the reflectivity of the cavity front mirror. Furthermore, for fixed front-mirror reflectivity, adjusting the intrinsic absorption in the ultrathin layer (e.g., via an electric potential for graphene \cite{Kasry10ACSNano}) can maximize the net cavity absorption.

The notion of omniresonance is a subclass of the more general framework of \textit{classical entanglement}, that refers to classical optical fields in which non-separability is created by introducing deterministic correlations between physically independent degrees of freedom \cite{Spreeuw98FoundPhys,Luis09OptCommun,Qian11OL,Kagalwala13NP,Toppel14NJP,Berg-Johansen15Optica}. Here, classical entanglement is introduced between two continuous degrees of freedom: the wavelengths and their associated angles of incidence, whereby each wavelength is assigned to a single prescribed angle \cite{Kondakci16OE}, resulting thereby in omni-resonance. In addition to omni-resonance, a similar conception of classical entanglement has been instrumental in developing a new class of propagation-invariant pulsed beams of wave packets that we have coined the name `space-time' (ST) wave packets \cite{Kondakci17NatPhotonics,Yessenov19PRA,Kondakci19NatCommun,Shiri20Arxiv}. Moreover, we have recently established the connection between omni-resonance and ST wave packets and have demonstrated that a family of such pulsed beams are omni-resonant and can thus traverse planar FP cavities without change even when the pulse bandwidth exceeds the cavity resonant linewidth \cite{Shiri20OL, Shiri20APLP}.

The effect demonstrated here is an important step towards the utilization of CPA in energy harvesting technologies despite the inherently resonant nature of CPA. It is an open question whether the free-space arrangement we made use of here (a diffraction grating and a lens) can be replaced by a single element (such as a metasurface) that may be integrated with the multilayer cavity structure into a single device. This may pave the way to future highly-efficient ultrathin solar cell devices and photodetectors.

\begin{acknowledgments}
This work was funded by the U.S. Air Force Office of Scientific Research (AFOSR) under MURI award A9550-14-1-0037, and by the U.S. Office of Naval Research (ONR) under MURI award N00014-20-1-2789.
\end{acknowledgments}

\appendix
\section{FP cavity design and construction}\label{App:MirrorDesign}

The Bragg mirrors used in constructing the FP resonator comprise bilayers of $94$-nm-thick SiO$_2$ ($n\!=\!1.46$ at $\lambda\!=\!550$~nm) and $61$-nm-thick Ti$_2$O$_3$ ($n\!=\!2.3$ at $\lambda\!=\!550$~nm) deposited via e-beam evaporation. The Ti$_2$O$_3$ films were formed by evaporating TiO$_2$ source material under O$_2$ partial pressure. The cavity (total thickness $\approx\!6.4$~$\mu$m) is deposited monolithically on a $1$-mm-thick, $25$-mm-diameter glass slide. The layered structure from the top (the incidence port on M$_1$) is as following: Air–(LH)$_5$–[SiO$_2$/Graphene/SiO$_2$]–(LH)$_{10}$–glass. The thickness of the two SiO$_2$ layers are $\approx\!0.1~ \mu m$ (adjucent to M$_1$) and $\approx\!4~ \mu m$ (adjucent to M$_2$). Each bilayer (LH) consists of low-index SiO$_2$ (L) and high-index Ti$_2$O$_3$ (H) material. The result is a $100$-nm-wide spectral reflection band with $\sim94$\% reflectivity at the center wavelength $\lambda_\mathrm{c}\!\approx\!550$ nm at normal incidence from air ($\sim96\%$ at normal incidence from SiO$_2$). 

\section{Omni-resonance pre-conditioning arrangement}
\label{App:SetupDesign}

\begin{figure}[t!]\label{Fig:Preconditioning}
\centering\includegraphics[width=8.6cm]{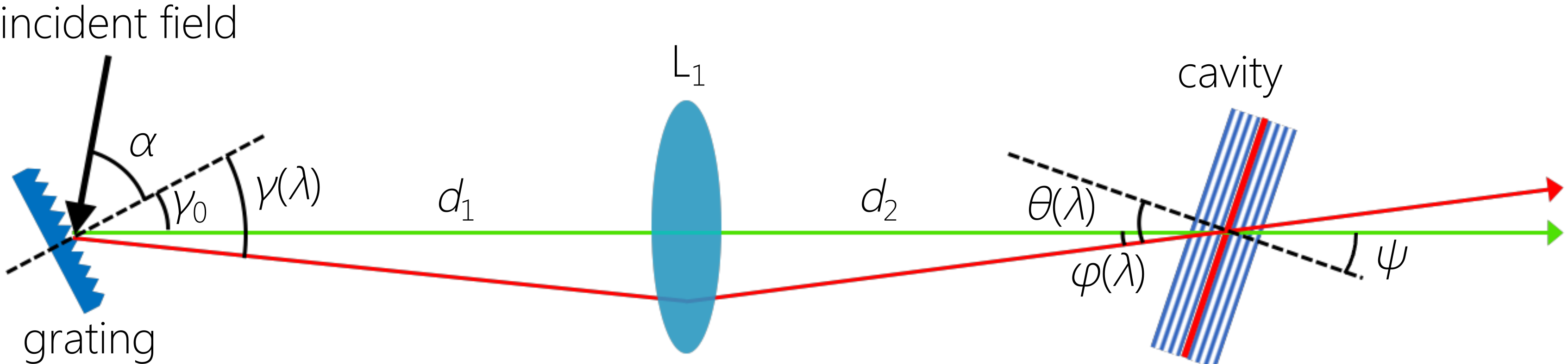}
\caption{Schematic depiction of the pre-conditioning optical system the introduces the necessary angular dispersion for realizing omni-resonance. The configuration comprises a diffraction grating and and a lens L$_1$; $\alpha$ and $\gamma(\lambda)$ are measured with respect to the grating normal. The optical axis (shown in green) coincides with $\gamma_\mathrm{o}\!=\!\gamma(\lambda_\mathrm{c}\!=\!550$~nm). Light is then incident on the cavity, which is tilted an angle $\psi$ with respect to the optical axis; $\varphi(\lambda)$ is measured with respect to the optical axis, while $\theta(\lambda)$ is measured from the cavity normal: $\theta(\lambda)\!=\!\varphi(\lambda)+\psi$.}
\end{figure}

A detailed configuration for the pre-conditioning system that modulates the incident optical field before impinging on the FP cavity is depicted in Fig.~\ref{Fig:Preconditioning}; see also \cite{Shabahang17SciRep,Shabahang19OL,Villinger20arxiv}. We assume an ideal grating, with linearly polarization collimated white light from the halogen lamp directed at an incidence angle $\alpha\!=\!50^\circ$ with respect to the grating normal. The angularly dispersed light from the grating is directed to the FP cavity via a lens L$_1$. We take $\lambda\!=\!550$~nm as the central wavelength that defines the optical axis. The tilt angle $\psi$ of the cavity is measured with respect to this optical axis. We define the angle $\gamma(\lambda)$, which is the diffraction angle of wavelength $\lambda$ with respect to the grating normal. The central wavelength $\lambda_\mathrm{c}\!=\!550$ nm is diffracted at $\gamma(\lambda_\mathrm{c}\!=\!550$ nm)$=\gamma_\mathrm{o}$ and coincides with the optical axis, and any other wavelength $\lambda$ makes an angle $\gamma(\lambda)\!-\!\gamma_\mathrm{o}$ with respect to the optical axis. This angle is \textit{boosted} via the lens L$_1$ by a ratio $d_1/d_2$ , where $d_1$ and $d_2$ are the distances from the grating to L$_1$ and from L$_1$ to the cavity, respectively. The incidence angle with respect to the optical axis after L$_1$ is 
\begin{equation}
\varphi(\lambda)=\tan^{-1}\left\{\frac{d_1}{d_2}\tan[\gamma(\lambda)\!-\!\gamma_\mathrm{o}]\right\},
\end{equation}
with $\varphi_\mathrm{o}\!=\!\varphi(\lambda_\mathrm{c}\!=\!550 \mathrm{nm})\!=\!0$. The distances $d_1$ and $d_2$ are selected such that the illuminated spot on the grating is imaged onto the cavity. If the focal length of L$_1$ is $f$, then $d_2\!=\!\frac{fd_1}{f-d_1}$. When the cavity is oriented such that it is perpendicular to the optical axis, the angle of incidence of each wavelength is $\varphi(\lambda)$. Upon tilting the cavity by $\psi$, the angle of incidence with respect to the normal to the cavity is $\theta(\lambda)\!=\!\varphi(\lambda)+\psi$. By adjusting $\psi$, one can provide the optimal angular dispersion required to achieve the omni-resonance condition.

\section{transfer matrix analysis for the CPA cavity}
\label{App:TMM}

The scattering matrices associated with the front and back mirrors ($\mathbf{S}_1$ and $\mathbf{S}_2$, respectively) take the form:
\begin{equation}
\mathbf{S}_j\!=\!\begin{pmatrix} t_{j}e^{i\beta_{j}} & r_{j}e^{i\alpha_{j}} \\ -r_{j}e^{i(2\beta_{j}\!-\!\alpha_{j})} & t_{j}e^{i\beta_{j}}\end{pmatrix},
\end{equation}
where $t_{j}$ and $r_{j}$ ($j\!=\!1,2$) are the transmission and reflection amplitudes of the mirrors, respectively, and $\beta_{j}$ and $\alpha_{j}$ are the associated phases. The scattering matrix relates the right and left outgoing waves ($A_\mathrm{R}$ and $B_\mathrm{L}$, respectively) to the left and right incoming waves ($A_\mathrm{L}$ and $B_\mathrm{R}$, respectively). The transfer matrix $\mathbf{M}$ corresponding to $\mathbf{S}$ is:
\begin{equation}\label{Eq:MirrorTransferMatrix}
\mathbf{M}_{j}=\frac{1}{t_{j}}\begin{pmatrix} e^{-i\beta_{j}} & -r_{j} e^{-i(\beta_{j}\!-\!\alpha_{j})} \\ -r_{j} e^{i(\beta_{j}\!-\!\alpha_{j})} & e^{i\beta_{j}}\end{pmatrix}.
\end{equation}

Propagation for a distance $d$ in a medium of refractive index $n$ is represented by a transfer matrix:
\begin{equation}\label{Eq:PropTransferMatrix}
\mathbf{M}(\varphi)=\begin{pmatrix} e^{-i\varphi} & 0 \\ 0 & e^{i\varphi} \end{pmatrix}
\end{equation}
where $\varphi\!=\!nkd$ is the optical phase. This transfer matrix applies to propagation for a distance $d$ from M$_1$ to the graphene monolayer, and then a distance $L-d$ to M$_2$.

\bibliography{Refs}% Produces the bibliography via BibTeX.

\end{document}